\documentclass[aps,prl,twocolumn,superscriptaddress]{revtex4}

\usepackage{amsmath,amssymb}
\usepackage{graphicx}
\usepackage{textcomp}
\usepackage{units}
\usepackage{mciteplus}

\begin{document}

\title{\textbf{Salt-induced changes of colloidal interactions in critical mixtures}}

\author{Ursula Nellen}
\affiliation{2. Physikalisches Institut, Universit\"at Stuttgart,
Pfaffenwaldring 57, 70550 Stuttgart, Germany}

\author{Julian Dietrich}
\affiliation{2. Physikalisches Institut, Universit\"at Stuttgart,
Pfaffenwaldring 57, 70550 Stuttgart, Germany}

\author{Laurent Helden}
\affiliation{2. Physikalisches Institut, Universit\"at Stuttgart,
Pfaffenwaldring 57, 70550 Stuttgart, Germany}

\author{Shirish Chodankar}
\affiliation{Paul Scherrer Institut, CH-5232 Villigen PSI, Switzerland. }
\affiliation{$^{c}$~ETH Z\"{u}rich, CH-8093 Z\"{u}rich, Switzerland. }

\author{Kim Nyg{\aa}rd}
\affiliation{Paul Scherrer Institut, CH-5232 Villigen PSI, Switzerland. }
\affiliation{$^{c}$~ETH Z\"{u}rich, CH-8093 Z\"{u}rich, Switzerland. }

\author{J. Friso van der Veen}
\affiliation{Paul Scherrer Institut, CH-5232 Villigen PSI, Switzerland. }
\affiliation{$^{c}$~ETH Z\"{u}rich, CH-8093 Z\"{u}rich, Switzerland. }

\author{Clemens Bechinger}
\affiliation{2. Physikalisches Institut, Universit\"at Stuttgart,
Pfaffenwaldring 57, 70550 Stuttgart, Germany}
\affiliation{Max-Planck-Institut f\"{u}r Intelligente Systeme, Heisenbergstra{\ss}e 3, 70569 Stuttgart, Germany.}

\begin{abstract}
\vspace{0.4cm}
We report on salt-dependent interaction potentials of a single charged particle suspended in a binary liquid mixture above a charged wall. For symmetric boundary conditions (BC) we observe attractive particle-wall interaction forces which are similar to critical Casimir forces previously observed in salt-free mixtures. However, in case of antisymmetric BC we find a temperature-dependent crossover from attractive to repulsive forces which is in strong contrast to salt-free conditions. Additionally performed small-angle x-ray scattering experiments demonstrate that the bulk critical fluctuations are not affected by the addition of salt. This suggests that the observed crossover can not be attributed alone to critical Casimir forces. Instead our experiments point towards a possible coupling between the ionic distributions and the concentration profiles in the binary mixture which then affects the interaction potentials in such systems.
\end{abstract}

\maketitle

\section{Introduction}
Critical Casimir forces currently receive considerable attention both from a fundamental point of view but also as a versatile {\em in situ} mechanism for changing pair potentials in colloidal systems \cite{fis78,her08,lu10,sch03, nel09,soy08, gam09, bon09, *bon09b, *bon09c}. Such forces are induced by the spatial confinement of critical fluctuations of the concentration in a binary liquid mixture near its critical point. The spatial extension of such fluctuations which also sets the range of critical Casimir forces is given by the bulk correlation length $\xi$ of the mixture which diverges upon approaching the critical point. Because $\xi$ strongly depends on the temperature $T$, critical Casimir forces respond to minute temperature changes close to its critical value $T_C$. In addition, they can be changed from attractive to repulsive by small chemical modifications of the particles surfaces which alter their adsorption preference
for the mixture's components \cite{soy08,nel09}. Accordingly, critical Casimir forces provide an interesting possibility to reversibly tune pair interactions in colloidal systems. Recent experiments quantitatively confirmed the theoretically predicted dependence of such forces on both, the mixture's temperature and the surface properties (BC) of the colloidal particles \cite{her08, nel09, gam09}.\\
Critical Casimir forces also sensitively depend on other parameters such as electric field gradients. Experiments have demonstrated a measurable shift of the critical point in the presence of unevenly spaced electrodes \cite{tso04}. Alternatively, electric field gradients are also obtained by the addition of ions to critical mixtures. Due to differences in the solubility of the salt ions in the mixture's components this leads to salt-dependent changes of the concentration profiles \cite{onu04, sad09,  cia10, ben09} and the bulk phase behavior of binary mixtures \cite{sea93,bal99}. Furthermore, salt can also modify the surface adsorption preference as confirmed in capillary rise \cite{sig86} and light scattering \cite{dui91} experiments.
Here, we address the question how salt ions change the interaction between a particle and a wall suspended in a critical binary mixture. We observe that even at salt concentrations for which no changes in the bulk phase behavior are observed, the interaction between a single particle and a wall in a critical binary mixture is largely changed compared to salt-free conditions.

\section{Experimental System}
Our experimental system consists of a binary solvent composed of water and 2,6-lutidine. The phase diagram of such mixtures has a lower critical point at $T_C=304 \pm 0.1\,\mathrm{K}$ and a lutidine mass fraction of $c_L=0.286 \pm 0.05$. All measurements shown in this paper were performed below $T_C$. We used 10~mM of potassium bromide (KBr) to investigate the influence of additional salt on such mixtures because KBr only weakly shifts the critical point \cite{sea93}. The absence of a salt-dependent shift of the critical point has also been confirmed by our own measurements using the equal volume method \cite{val10}.
Due to the difference in the static dielectric constants of water $\epsilon_W=79.5$ and lutidine $\epsilon_L= 6.9$ (data were taken at 298.15~K \cite{lan08}) the solubility of KBr in water and lutidine differs drastically.
The solubility of KBr in water is well known: 39.39\% (mass percentage, at 298~K) \cite{wan08}. In order to determine the solubility of KBr in lutidine, we prepared an oversaturated lutidine-KBr mixture. After equilibration part of the liquid was extracted and evaporated to measure the weight of the remaining dry salt. In our experiments we did not find any indication that KBr is dissolved in lutidine within our experimental accuracy which corresponds to an upper limit of the KBr solubility of 0.03\% .

\section{Results}

\subsection*{Small-Angle X-Ray Scattering}
\begin{figure}
	\centering
	\includegraphics[width=0.48\textwidth]{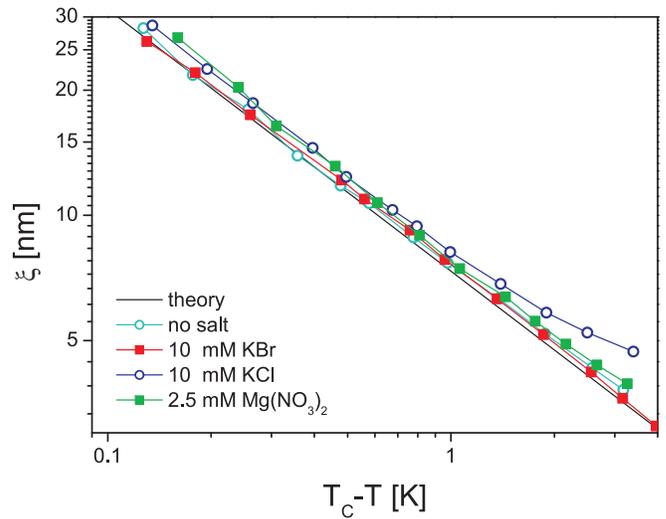}
	\caption{Temperature dependence of the correlation length in a water 2,6-lutidine mixture for different types and concentrations of ions. The black line shows the theoretically predicted Ising like behavior for $\xi_{0}=0.2$ nm and $\nu=0.63$.}
	\label{fig3}
\end{figure}

The bulk properties of the binary liquid in the presence of salt ions were investigated by small-angle x-ray scattering (SAXS) experiments at the coherent SAXS beamline (cSAXS) of the Swiss Light Source at the Paul Scherrer Institut. The incident x-ray radiation (wavelength $\lambda$=1.00 \AA), impinging normal to the symmetry axis of the cylindrical sample cell, was focused onto the detector plane in order to increase the angular resolution. The scattered x-rays were collected using a two-dimensional, single-photon-counting pixel detector (PILATUS 2M, pixel size 172$\times$172~ $\mu$m$^2$ with a total of 1461$\times$1560 pixels \cite{kra09}), which was positioned 7~m behind the sample. An evacuated flight tube between the sample and the detector was employed in order to reduce the background scattering.
>From the measured scattering intensity $I(q)$ as a function of momentum transfer $q$ we obtained after subtraction of a temperature-independent background the correlation length $\xi$ using the Ornstein-Zernike relation:\cite{han06}
\begin{equation}
I\left(q\right) \propto \frac{1}{1+q^2\xi^2}.
\label{equOrn}
\end{equation}
The low contrast between water and lutidine and the parasitic background limited the determination of the correlation length between 3 - 30~nm.
Fig. \ref{fig3} shows the temperature-dependent correlation length of a pure critical water-lutidine mixture and one with 10~mM KBr in a log-log plot. Obviously, both measurements coincide which demonstrates that the concentration fluctuations  of the mixture are not modified by the salt ions on the probed length scale. In addition, the measured temperature dependence of the correlation length $\xi(T)$ is in quantitative agreement with predictions obtained within the three-dimensional Ising-type universality class. The solid line in Fig. \ref{fig3} is given by
\begin{equation}
\xi = \xi_{0} \left(1-\frac{T}{T_C}\right)^{-\nu}
\label{equcorr}
\end{equation}
with the critical exponent $\nu=0.63$ \cite{gul72} and the amplitude $\xi_{0}$ set by the typical range of the molecular pair potential in the liquid. Independent measurements in critical water-lutidine mixtures determined this value to $\xi_{0}=0.2$ nm \cite{gul72}. For comparison, we also measured the correlation length in water-lutidine mixtures for other types of salt (potassium chloride (KCl) and magnesiumnitrate ($\mathrm{Mg(NO_3)_2}$). The corresponding values of $\xi_0$ and $\nu$ for which best agreement between the measured $\xi (T)$ and Eqn. \ref{equcorr} was obtained are shown in table \ref{table:test}. The salt concentrations have been chosen to yield the same Debye screening length of 2.6~nm for each solution. From those measurements it is obvious that addition of salt does not influence the bulk properties of the liquid mixture close to the critical point.
\begin{table}[h]
\small
	\begin{tabular*}{0.48\textwidth}{@{\extracolsep{\fill}}lll}
	\hline
	Salt&$~\xi_0$~[nm]&$~\nu$\\
	\hline
	no salt &~0.20 &~0.63\\
	10 ~mM KBr&~0.24 &~0.61\\
	10 ~mM KCl &~0.25 &~0.61\\
	2.5~mM $\mathrm{Mg(NO_3)_2}$ &~0.18 &~0.66 \\
		\hline
	\hline
	\end{tabular*}
\caption{Fitting parameters for the SAXS data.}
\label{table:test}
\end{table}

\subsection*{Total Internal Reflection Microscopy}
Interaction potentials between a negatively charged polystyrene (PS) sphere with radius $a=0.75\,\mathrm{ \mu m}$ and a glass surface were determined in a critical water-lutidine mixture with total internal reflection microscopy (TIRM). This method is based on the measurement of the vertical Brownian motion of a single colloidal particle by means of scattered light in the presence of an evanescent field. The evanescent field is created by total internal reflection of a laser beam impinging at a specific angle at the interface between a glass prism and the binary mixture. When the relationship between the intensity of the scattered light $I$ and the distance of the particle from the wall $z$  is known, the particle-wall interaction potential can be determined with an accuracy of approximately 0.1 $k_BT$  \cite{pri90, wal97}. In case the function $I(z)$ is not known {\it a priori}, it can be experimentally determined by a procedure described in \cite{vol09}. From these measurements we obtained under our experimental conditions  $I(z)\propto \exp(- \beta \cdot z )$ with $\beta^{-1} = 138\pm4$\,nm the penetration depth of the evanescent field. This relationship was found to be valid over the entire temperature range and for all boundary conditions presented in this work. Due to a small temperature dependence of the refractive index of the water-lutidine mixture ($\Delta n\leq 2.4\cdot 10^{-3}$ for $T_C-T\leq 10$\,K) the penetration depth slightly varies with temperature. For the temperature range considered here, however, the corresponding effects on the measured potentials can be neglected within the experimental resolution of TIRM. The boundary condtions are given by the preferential adsorption behavior of the mixture's components which can be altered by appropriate functionalization of the corresponding surfaces \cite{nel09}. The glass surface was rendered hydrophobic (+~BC) by silanization with hexamethyldisilazane or hydrophilic (-~BC) by exposure to an oxygen plasma for about a minute. Because the PS particle is hydrophilic, this allowed us to compare symmetric (-~-) and antisymmetric (-~+) BC.
\begin{figure}
\centering
	\includegraphics[width=0.48\textwidth]{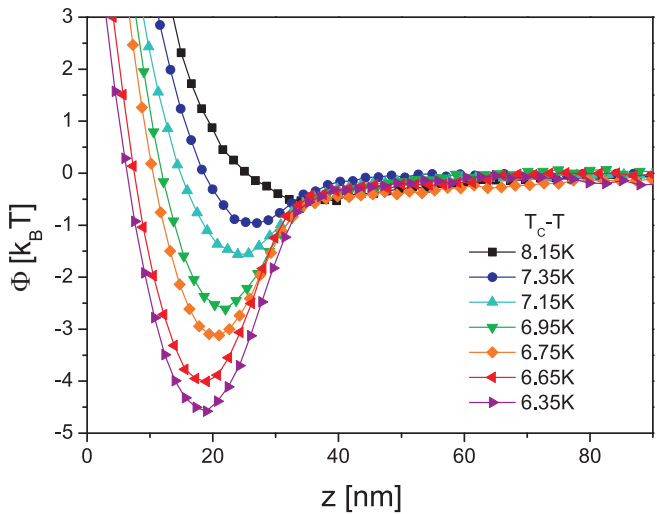}
	\caption{Temperature-dependent potentials between a PS sphere ($R=0.75~\mu$m) and a wall in a critical system of water and 2,6-lutidine in the presence of 10~mM KBr for symmetric BC.}
	\label{fig1}
\end{figure}
Fig.\,\ref{fig1} shows the interaction potential between a PS sphere and a silica wall in a critical mixture with 10~mM KBr for symmetric (-~-) BC where critical Casimir forces are expected to be attractive. It should be noted that similar potentials have also been observed for other salt concentrations. Far below the critical temperature ($T_C - T > 8$ K), no indication for critical Casimir forces can be observed and the sphere-wall potentials are only composed of a temperature-independent short-ranged electrostatic repulsion. The weight of the PS particle adds a linear contribution to the potentials which does not change and is subtracted in all measurements \cite{her08}. In contrast to salt-free conditions for which attractive critical Casimir forces were only observed in the immediate vicinity of $T_C$ ($T_C - T<1$~K) \cite{her08}, here an attractive force appears already 7.35~K below $T_C$. Increasing the temperature even further the potential well becomes too deep to be measured with TIRM. However, since the particle remains localised very close to the wall at higher temperatures, we conclude that the forces stay attractive up to the critical point. Such an increase of the amplitude of critical Casimir forces by addition of ions is consistent with recent observations of other authors \cite{bon09, *bon09b, *bon09c} and may be explained by the reduced particle-wall distance $z$ due to screening of surface charges. The critical Casimir interaction \cite{her08}
\begin{equation}
\frac{\Phi_C}{k_{B}T}=\frac{a}{z}\, \theta(z/\xi)
\label{casipot}
\end{equation}
is proportional to $1/z$ and the universal scaling function $\theta(z/\xi)$, which increases monotonically in absolute value with decreasing $z/\xi$ in the accessed range.
 Accordingly, for a certain value of $\xi$, the critical Casimir contribution turns out to be significant at small distances (as sampled here) while negligible at larger distances (such as the ones sampled in previous experiments \cite{her08}) and thus explains why such forces already appear far below $T_C$ in the presence of salt.
\begin{figure}
\centering
	\includegraphics[width=0.48\textwidth]{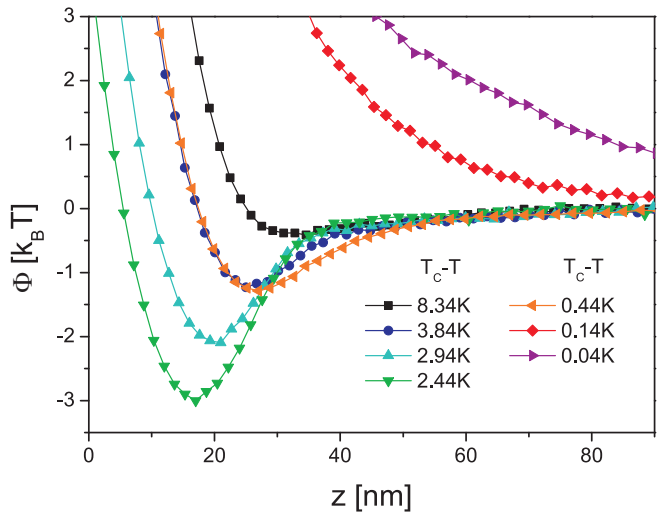}
	\caption{Temperature-dependent particle-wall potentials resulting for antisymmetric BC. The potentials were probed by a PS colloid suspended in a critical water and 2,6-lutidine mixture containing 10~mM KBr.}
	\label{fig2}
\end{figure}
Fig. \ref{fig2} shows the corresponding particle-wall potentials for antisymmetric (-~+) BC. Similar as above, a strongly temperature-dependent force occurs several Kelvin below $T_C$. However, in
contrast to salt-free conditions where the critical Casimir forces appearing close to $T_C$  are repulsive for antisymmetric BC, here we find an attraction up to $T_C - T=2.44$~K. Increasing the temperature even further the particle becomes increasingly localized close to the wall (as inferred from their light scattering) which demonstrates that the attraction becomes even larger with increasing temperature. Gratifyingly, very close to $T_C$ ($T_C - T \leq 0.44$ K) the attraction weakens and eventually the forces become repulsive, which is in line with the signature of critical Casimir forces. The same qualitative effect was observed for other salt concentrations, e.\,g. 5~mM. This behaviour, in particular the observed crossover from attractive to repulsive forces, is in contrast with previous salt-free measurements and clearly emphasizes the importance of salt for such interactions.

\subsection*{Surface Plasmon Resonance}
\begin{figure}
\centering
	\includegraphics[width=0.48\textwidth]{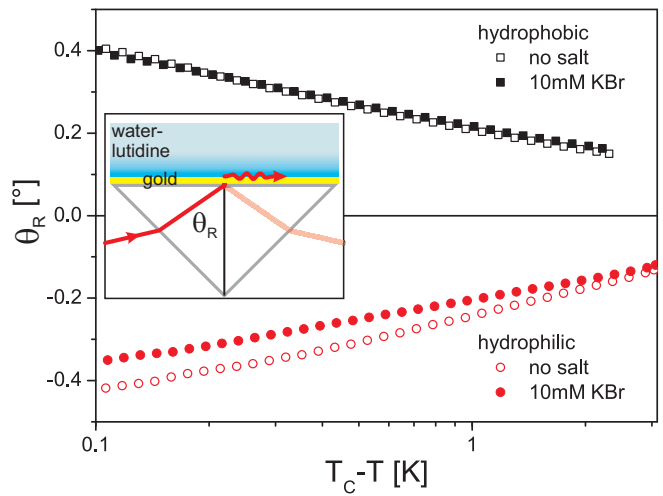}
	\caption{Temperature dependence of the SPR angle for hydrophilic (circles) and hydrophobic (squares) surface preference in a mixture of water and 2,6-lutidine with (full symbols) and without (open symbols) addition of 10~mM KBr. Inset: schematic SPR setup with prism-coupler.}
	\label{fig4}
	\end{figure}

As mentioned above, the sign of critical Casimir forces is determined by the BC of the involved surfaces which are determined by their adsorption preference for one of the two components of the binary liquid mixture. The concentration of this preferred component decays perpendicular to the surface on a length scale set by the bulk correlation length $\xi(T)$ of the mixture \cite{liu89,flo95,law01}. In order to decide whether the above crossover is a result of salt-induced changes of the BC, we investigated the temperature-dependence of the adsorption profile of the mixture on hydrophilic and hydrophobic surfaces with a surface plasmon resonance experiment (SPR). This was achieved by the excitation of surface plasmons at a metal/dielectric interface by attenuated total reflection. The dispersion relation of surface plasmons is rather sensitive to changes in the permittivity of the dielectric medium, which results in a change of the angle of incidence for which the resonance condition is met and the laser beam is not reflected\cite{rae88,kno98}. Due to the contrast in the refractive indices of water ($n_W=1.33$) and lutidine ($n_L=1.49$), a change in the adsorption profile generates a different dielectric environment resulting in a change of the resonant incidence angle $\theta_R$. Although these measurements do not allow for spatially resolved adsorption profiles, they distinguish whether water or lutidine is the preferred component at the corresponding surface.
We used a high refractive index glass prism (Schott N-LASF9) with a 50 nm thin layer of gold and a Coherent Radius laser ($\lambda = 635$ nm) in a typical Kretschmann configuration \cite{rae88}, depicted in the inset of Fig. \ref{fig4}. The gold surface was either treated with 1-octadecanethiol or 11-hydroxy-1-undecanethiol to render it hydrophobic or hydrophilic providing identical functional endgroups as those used in the TIRM experiments.
Fig. \ref{fig4} shows the measured temperature dependence of the shift of the angle of resonance $\Delta\theta_R$ for a critical water-lutidine mixture with and without 10~mM KBr.Positive and negative values indicate an enrichment of lutidine and water, respectively. Independent of the salt concentration the shift monotonically increases (decreases) for hydrophobic (hydrophilic) surfaces upon approaching $T_C$ which reflects the increasing enrichment of the corresponding component at the surface. Although the addition of KBr slightly changes the adsorption profile on the hydrophilic surface, the general adsorption preference (i.e. BC) is not changed and thus does not provide an explanation for the observed crossover.

\section{Discussion and Conclusion}
Our measurements demonstrate that colloidal interactions in critical mixtures are strongly modified by the presence of salt and thus provide novel opportunities to tune the pair interactions in such systems. Contrary to salt-free conditions, where critical Casimir forces occur only very close to the critical point, the addition of salt enhances electrostatic screening and thus reduces typical particle-wall distances leading to enhanced critical Casimir forces which already set in several Kelvin away from $T_C$. In case of symmetric BC the behaviour can be qualitatively explained by the reduced particle-wall (or colloid-colloid) distances which in turn lead to stronger critical Casimir attractions. More interestingly, for antisymmetric BC we find a crossover from attractive to repulsive forces when approaching the critical point. Because the correlation length and the effective BC are not altered by the salt as confirmed by SAXS and SPR experiments, this suggests that the observed crossover can not be attributed alone to critical Casimir forces. Instead, we suppose that due to the strong differences in solubility of KBr in water and lutidine, the ionic profiles may be strongly coupled to the concentration profiles of the mixture which in turn may lead to interesting changes in electrostatic interactions. Because colloidal interactions in binary mixtures near the critical point are also important for the formation of bijels or glass formation \cite{thi10,lu10}, we expect that our findings are important for other systems as well.

\section{Acknowledgements}
We thank M.~Bier, A.~Maciolek,  A.~Gambassi, S.~Dietrich, S.~Samin and Y.~Tsori for stimulating and helpful discussions and A.~Menzel for assistance during the SAXS experiment.
We also acknowledge funding from the Deutsche Forschungsgemeinschaft BE 1788/9 and the European Community's Seventh Framework Programme (FP7/2007-2013) under grant agreement $\mathrm{n}^\circ$ 226716.

\bibliography{papers}
\bibliographystyle{rsc}

\end{document}